\documentclass[a4paper, reprint, amsfonts, amssymb, amsmath]{article}

\usepackage{amsfonts}
\usepackage{amssymb}
\usepackage{amsmath}
\usepackage{bm}
\usepackage[utf8]{inputenc}
\usepackage{mathtools}
\usepackage{authblk}

\usepackage[english, english]{babel}
\usepackage{xfrac}
\usepackage{braket}
\usepackage[section]{placeins}
\usepackage{wrapfig}
\usepackage{graphicx}

\begin{document}
  \selectlanguage{english}
  \title{Correction to the quantum phase operator for photons}
  \date{\fontsize{10}{15}\selectfont \today}
  \author{Mads J. Damgaard}
  \affil{\fontsize{10}{15}\selectfont\textit{Niels Bohr Institute, University of Copenhagen, 2100 Copenhagen, Denmark}}
  \maketitle

  \begin{abstract}
  The vector potential operator, $\hat{\boldsymbol A}$, is transformed and rewritten in terms of cosine and sine functions in order to get a clear picture of how the photon states relate to the $\boldsymbol A$ field. The phase operator, defined by $\hat E = \exp(-i \hat \phi)$, is derived from this picture. The result has a close resemblance with the known Susskind-Glogower (SG) operator, which is given by $\hat E_{SG}=(\hat a_{\boldsymbol k} \hat a_{\boldsymbol k}^\dagger)^{-1/2} \hat a_{\boldsymbol k}$. It will be shown that $\hat a_{\boldsymbol k}$ should be replaced by $(\hat a_{\boldsymbol k} + \hat a_{-\boldsymbol k}^\dagger)$ instead to yield $\hat E = ((\hat a_{\boldsymbol k} + \hat a_{-\boldsymbol k}^\dagger ) (\hat a_{\boldsymbol k}^\dagger + \hat a_{-\boldsymbol k}))^{-1/2} (\hat a_{\boldsymbol k} + \hat a_{-\boldsymbol k}^\dagger)$, which makes the operator unitary. $\hat E$ will also be analyzed when restricted to the space of only forward moving photons with wave vector $\boldsymbol k$. The resulting phase operator, $\hat E_+$, will turn out to resemble the SG operator as well, but with a small correction: Whereas $E_{SG}$ can be equivalently written as $\hat E_{SG} =  \sum_{n=0}^{\infty} |n\rangle \langle n+1 |$, the operator, $\hat E_+$, is instead given by $\hat E_+ = \sum_{n=0}^{\infty} a_n |n \rangle \langle n+1|$, where $a_n = (n+1/2)!/(n! \sqrt{n+1})$. The sequence, $(a_n)_{n \in \lbrace 0, 1, 2, \ldots \rbrace}$, converges to $1$ from below for $n$ going to infinity.
  \end{abstract}

  \section*{Introduction}
  Electromagnetic waves are classically described by a wave vector, an amplitude and a phase. In quantum mechanics the concept of the wave vector remains the same, but the amplitude and the phase should now be represented by operators. Whereas there are general agreement about the amplitude operator, there has so far, according to Gerry and Knight \cite{Gerry and Knight}, not been found a phase operator, that satisfies all the expected requirements. One of the quite successful proposals according to Gerry and Knight \cite{Gerry and Knight} is the Susskind-Glogower (SG) operator, which is given by
  \begin{equation}
  \hat E_{SG} =  (\hat a \hat a^\dagger)^{-\frac{1}{2}} \hat a,
  \label{SG}
  \end{equation}
  or alternatively by
  \begin{equation}
  \hat E_{SG} =  \sum_{n=0}^{\infty} |n\rangle \langle n+1 |.
  \label{SG2}
  \end{equation}
  Here, $\hat a^\dagger$ and $\hat a$ are the creation and annihilation operators for photons with an implicit wave vector, $\bm k$, and $\ket{n}$ is the state with $n$ of these photons. The SG operator is not supposed to represent the phase, $\phi$, but the phase factor, $e^{i \phi}$. This means the SG operator is supposed to be unitary, but since
  \begin{equation}
  \hat E_{SG}^\dagger \hat E_{SG} = \hat I - |0 \rangle\langle 0| \neq \hat I,
  \end{equation}
  this is not the case.
  
  When I wrote my bachelor thesis \cite{bach_thes}, in which I looked at QED, my approach to the subject led me to quantize the electromagnetic field in terms of the cosine and sine Fourier components. This turned out to give a very neat picture of the quantum field with a quite easy interpretation of the photons, and where a unitary phase operator was easily derived.
  
  In this paper I will summarize the resulting phase operator of my bachelor thesis \cite{bach_thes}, but instead of starting by considering the electromagnetic field and how to quantize it, I will obtain the same operator simply by deriving it from the well known operator, $\hat{\bm A}$, representing the electromagnetic vector potential.

  \section*{Resolving $\hat{\bm A}$ in terms of cosine and sine functions}
  To derive the phase operator in this paper, we will start by considering the operator representing the electromagnetic vector potential\footnote{We could just as well have chosen to look at the operator for the electric field, $\hat{\bm E}(\bm x)$, instead since all of the following arguments apply for $\hat{\bm E}(\bm x)$ as well.}, $\hat{\bm A}(\bm x)$, expressed in terms of creation and annihilation operators, where $\bm x$ is a point in space.
  According to various textbooks, for example Gerry and Knight \cite{Gerry and Knight}, we have
  \begin{align}
  \hat{\bm A}(\bm x) =
  \sum_{\bm k, p} \frac{1}{\sqrt{2 k \mathcal{V}}}
  \big(\hat a_{\bm k, p} \, e^{i \bm k \cdot \bm x} +
  \hat a_{\bm k, p}^\dagger \, e^{-i \bm k \cdot \bm x} \big) \bm e_{\bm k, p},
  \label{A1}
  \end{align}
  where $\bm k$ is the photon wave vector, $p \in \lbrace 1, 2 \rbrace$ labels the two polarizations, $\bm e_{\bm k, 1}$ and $\bm e_{\bm k, 2}$ are two unit vectors orthogonal to $\bm k$ and to each other, $k = \sqrt{\bm k^2}$, and $\mathcal{V}$ is equal to the number of discrete values $\bm k$ runs through. The operators $\hat a_{\bm k, p}^\dagger$ and $\hat a_{\bm k, p}$ are the creation and annihilation operators, which respectively creates or annihilates a particle with wave vector $\bm k$ and polarization $p$. We  choose to look at the discretized case where $\bm k$ only take discrete values for simplicity.
  We will also need the commutation relations of the creation and annihilation operators, which are
  \begin{align}
  \begin{aligned}
  [\hat a_{\bm k, p}, \hat a_{\bm k', p}^\dagger] &= \delta_{k, k'}, \\
  [\hat a_{\bm k, p}, \hat a_{\bm k', p}] &= 0,
  \label{commute1}
  \end{aligned}
  \end{align}
  and we will need the Hamiltonian of the free photons, which is given by
  \begin{align}
  \begin{aligned}
  H = \sum_{\bm k, p} k 
  \Big(\hat a_{\bm k, p}^\dagger \hat a_{\bm k, p} +
  \frac{1}{2} \Big).
  \label{H1}
  \end{aligned}
  \end{align}
  
  The idea is now rewrite $\hat{\bm A}(\bm x)$ in terms of cosine and sine functions.
  This can be done by introducing the operators $\hat a_{cos, \bm k, p}$ and $\hat a_{sin, \bm k, p}$ defined by
  \begin{align}
  \hat a_{cos, \bm k, p} &= \frac{1}{\sqrt{2}}
  (\hat a_{\bm k, p} + \hat a_{-\bm k, p}),
  \label{a_cos} \\
  \hat a_{sin, \bm k, p} &= \frac{i}{\sqrt{2}}
  (\hat a_{\bm k, p} - \hat a_{-\bm k, p}).
  \label{a_sin}
  \end{align}
  These operators should only be defined for half of the $\bm k$-space, such that $\bm k \in \mathbb{R}^+ {\times} \mathbb{R}^2$. Let therefore $k_1 > 0$ in both $\hat a_{cos, \bm k, p}$ and $\hat a_{sin, \bm k, p}$, where $\bm k = (k_1, k_2, k_3)$. The inverted relations are
  \begin{align}
  \hat a_{\bm k, p} &= \frac{1}{\sqrt{2}}
  (\hat a_{cos, \bm k, p} - i \hat a_{sin, \bm k, p}),
  \label{a_d} \\
  \hat a_{-\bm k, p} &= \frac{1}{\sqrt{2}}
  (\hat a_{cos, \bm k, p} + i \hat a_{sin, \bm k, p}).
  \label{a_g}
  \end{align}
  Let us simplify the notation significantly in the following by suppressing some of the labels, except when we need to explicitly sum over them, and rewrite the operators as
  \begin{align}
  \hat a_{\bm k, p} \to \hat a_{\bm k}, \quad\quad 
  \hat a_{-\bm k, p} \to \hat a_{-\bm k}, \quad\quad 
  \hat a_{cos, \bm k, p} \to \hat a_{cos}, \quad\quad 
  \hat a_{sin, \bm k, p} \to \hat a_{sin}.
  \label{rename1}
  \end{align}
  
  The above transformation of the creation and annihilation operators allows us to rewrite the contribution to $\hat{\bm A}(\bm x)$ coming from $(\bm k, p)$ and $(-\bm k, p)$. Let us call this contribution $\hat{\bm A}_{\pm \bm k, p}(\bm x)$, such that
  \begin{align}
  \hat{\bm A}(\bm x) =
  \sum_{k_1 > 0, p} \hat{\bm A}_{\pm \bm k, p}(\bm x) 
  =\sum_{k_1 > 0, p} \frac{1}{\sqrt{2 k \mathcal{V}}}
  \big(\hat a_{\bm k, p} \, e^{i \bm k \cdot \bm x} + \hat a_{-\bm k, p} \, e^{-i \bm k \cdot \bm x}
  + \textrm{H.c.} \big) \bm e_{\bm k, p},
  \label{A_Apm}
  \end{align}
  where we have used the freedom in choosing the polarization vectors to set
  \begin{equation}
  	\bm e_{\bm k, p} = \bm e_{- \bm k, p}.
  \end{equation}
  Using eq.\ (\ref{a_d}) and (\ref{a_g}), and suppressing the labels according to prescription (\ref{rename1}), we can thus rewrite $\hat{\bm A}_{\pm \bm k, p}(\bm x)$ as
  \begin{align}
  \begin{aligned}
  \hat{\bm A}_{\pm \bm k, p}(\bm x)
  &= \frac{1}{\sqrt{2 k \mathcal{V}}}
  \big(\hat a_{\bm k} \, e^{i \bm k \cdot \bm x} + \hat a_{-\bm k} \, e^{-i \bm k \cdot \bm x}
  + \textrm{H.c.} \big) \bm e_{\bm k, p}  \\ &=
  \frac{1}{2 \sqrt{k \mathcal{V}}}
  \big((\hat a_{cos} - i \hat a_{sin}) \, e^{i \bm k \cdot \bm x} + (\hat a_{cos} + i \hat a_{sin}) \, e^{-i \bm k \cdot \bm x}
  + \textrm{H.c.} \big) \bm e_{\bm k, p}
  \\ &=
  \frac{1}{\sqrt{k \mathcal{V}}}
  \big((\hat a_{cos} + \hat a_{cos}^\dagger)\, \cos(\bm k \cdot \bm x) + (\hat a_{sin} + \hat a_{sin}^\dagger) \, \sin(\bm k \cdot \bm x) \big) \bm e_{\bm k, p}.
  \label{preA2}
  \end{aligned}
  \end{align}

  The next step is to identify the operators $(\hat a_{cos} + \hat a_{cos}^\dagger)/\sqrt{2 k}$ and $(\hat a_{sin} + \hat a_{sin}^\dagger)/\sqrt{2 k}$ as the ``position"\footnote{The quotation marks are used to remind the reader that the operator is not connected to the positions of the photons. It is instead connected with the amplitudes of the Fourier components of the $\bm A$ field.} operators in a two-dimensional harmonic oscillator system. To justify this, we need, firstly, to convince ourselves that $\hat a_{cos}$, $\hat a_{cos}^\dagger$, $\hat a_{sin}$ and $\hat a_{sin}^\dagger$ has the commutation relations of the ladder operators, i.e.\ that
  \begin{align}
  \begin{aligned}
  [\hat a_{cos}, \hat a_{cos}^\dagger] &= [\hat a_{sin}, \hat a_{sin}^\dagger] = 1, \\
  [\hat a_{cos}, \hat a_{sin}] &= 0.
  \label{commute2}
  \end{aligned}
  \end{align}
  It is easy to see that these relations indeed follow from the original commutation relations of eq.\ (\ref{commute1}). 
  Secondly, we need to show that the contribution to $H$ coming from $(\bm k, p)$ and $(-\bm k, p)$, given by
  \begin{align}
  \begin{aligned}
  H_{\pm \bm k, p} = k
  \big(\hat a_{\bm k}^\dagger \hat a_{\bm k} + \hat a_{-\bm k}^\dagger \hat a_{-\bm k} 
  + 1 \big),
  \label{H_ad}
  \end{aligned}
  \end{align}
  can be written equivalently as
  \begin{align}
  \begin{aligned}
  H_{\pm \bm k, p} = k \big(\hat a_{cos}^\dagger \hat a_{cos} + \hat a_{sin}^\dagger \hat a_{sin} 
  + 1 \big).
  \label{H2}
  \end{aligned}
  \end{align}
  To show this, we only need to use the following, very useful relations for the number operators, obtained from eq.\ (\ref{a_d}) and (\ref{a_g}):
  \begin{align}
  \hat a_{\bm k}^\dagger \hat a_{\bm k} = \frac{1}{2} \big(
  \hat a_{cos}^\dagger \hat a_{cos} +
  \hat a_{sin}^\dagger \hat a_{sin} -
  i \hat a_{cos}^\dagger \hat a_{sin} +
  i \hat a_{cos} \hat a_{sin}^\dagger
  \big),
  \label{N_d}
  \\
  \hat a_{-\bm k}^\dagger \hat a_{-\bm k} = \frac{1}{2} \big(
  \hat a_{cos}^\dagger \hat a_{cos} +
  \hat a_{sin}^\dagger \hat a_{sin} +
  i \hat a_{cos}^\dagger \hat a_{sin} -
  i \hat a_{cos} \hat a_{sin}^\dagger
  \big),
  \label{N_g}
  \end{align}
  and substitute them in eq.\ (\ref{H_ad}). This shows that $\hat a_{cos}$, $\hat a_{cos}^\dagger$, $\hat a_{sin}$ and $\hat a_{sin}^\dagger$ are indeed ladder operators in a two-dimensional harmonic oscillator system. Using the standard knowledge of harmonic oscillators, we can therefore rewrite $H_{\pm \bm k, p}$ in the more basic way in terms of Hermitian operators, $\hat A_{cos}$, $\hat A_{sin}$, $\hat p_{cos}$ and $\hat p_{sin}$, as
  \begin{align}
  \begin{aligned}
  H_{\pm \bm k, p} = \frac{1}{2}\hat p_{cos}^2 + \frac{1}{2}\hat p_{sin}^2 + \frac{1}{2} k^2 \hat A_{cos}^2 + \frac{1}{2} k^2 \hat A_{sin}^2,
  \label{H3}
  \end{aligned}
  \end{align}
  where
  \begin{align}
  \begin{aligned}
  \hat A_{cos} &= \frac{1}{\sqrt{2 k}} \big(\hat a_{cos} + \hat a_{cos}^\dagger \big), 
  \quad \hspace{0.37em} \hat A_{sin} = \frac{1}{\sqrt{2 k}} \big(\hat a_{sin} + \hat a_{sin}^\dagger \big), \\
  \hat p_{cos} &= \frac{-i \sqrt{k}}{\sqrt{2}} \big(\hat a_{cos} - \hat a_{cos}^\dagger \big), 
  \quad \hat p_{sin} = \frac{-i \sqrt{k}}{\sqrt{2}} \big(\hat a_{sin} - \hat a_{sin}^\dagger \big).
  \label{Ap_cossin}
  \end{aligned}
  \end{align}
  This can easily be checked using the commutation relations of eq.\ (\ref{commute2}).
  We can thus identify the two ``position" operators in eq.\ (\ref{preA2}) and finally rewrite $\hat{\bm A}_{\pm \bm k, p}(\bm x)$ as
\begin{align}
\begin{aligned}
  \hat{\bm A}_{\pm \bm k, p}(\bm x) = 
  \frac{\sqrt{2}}{\sqrt{\mathcal{V}}}
  \big(\hat A_{cos}  \cos(\bm k \cdot \bm x) + \hat A_{sin} \sin(\bm k \cdot \bm x) \big) \bm e_{\bm k, p}.
  \label{A2}
\end{aligned}
\end{align}

This concludes the transformation of $\hat{\bm A}_{\pm \bm k, p}(\bm x)$ and $H_{\pm \bm k, p}$. Since the coefficients of $\hat p_{cos}$ and $\hat p_{sin}$ vanish in eq.\ (\ref{A2}), the resulting quantum system is very easy to interpret, as the relation between the photon states and the physical $\bm A$ field is simple. To see this, let $\bm A_{\pm \bm k, p}$ be the contribution to $\bm A$ coming from the two Fourier coefficients, and let $A_{cos}$ and $A_{sin}$ be the observables associated with $\hat A_{cos}$ and $\hat A_{sin}$. If we then consider a wave function, $\psi$, defined on the $(A_{cos}, A_{sin})$-plane, we see from eq.\ (\ref{A2}) that each point, $(A_{cos}, A_{sin})$, in the plane corresponds to a physical configuration of the $\bm A_{\pm \bm k, p}$ field. 
The absolute square of the amplitude, $|\psi(A_{cos}, A_{sin})|^2$, thus represents the probability density of finding the amplitudes of the two Fourier components to be $(A_{cos}, A_{sin})$. Equation (\ref{H3}) furthermore tells us that the dynamics of $\psi$ are still simply those of a two-dimensional harmonic oscillator.

In the untransformed system where ``position" and ``momentum" operators are defined similarly to eq.\ (\ref{Ap_cossin}) but in terms of $\hat a_{\bm k}$ and $\hat a_{-\bm k}$ instead, the coefficients of the ``momentum" operators do not vanish in the expression for $\hat{\bm A}_{\pm \bm k, p}(\bm x)$. This is what makes the physical interpretation in this original system less immediate.

\begin{figure}[t]
	\centering
	\begin{minipage}[t]{0.40\textwidth}
		\includegraphics[width=1.0 \textwidth, trim={0cm, 0cm, 0cm, 0cm}, clip]{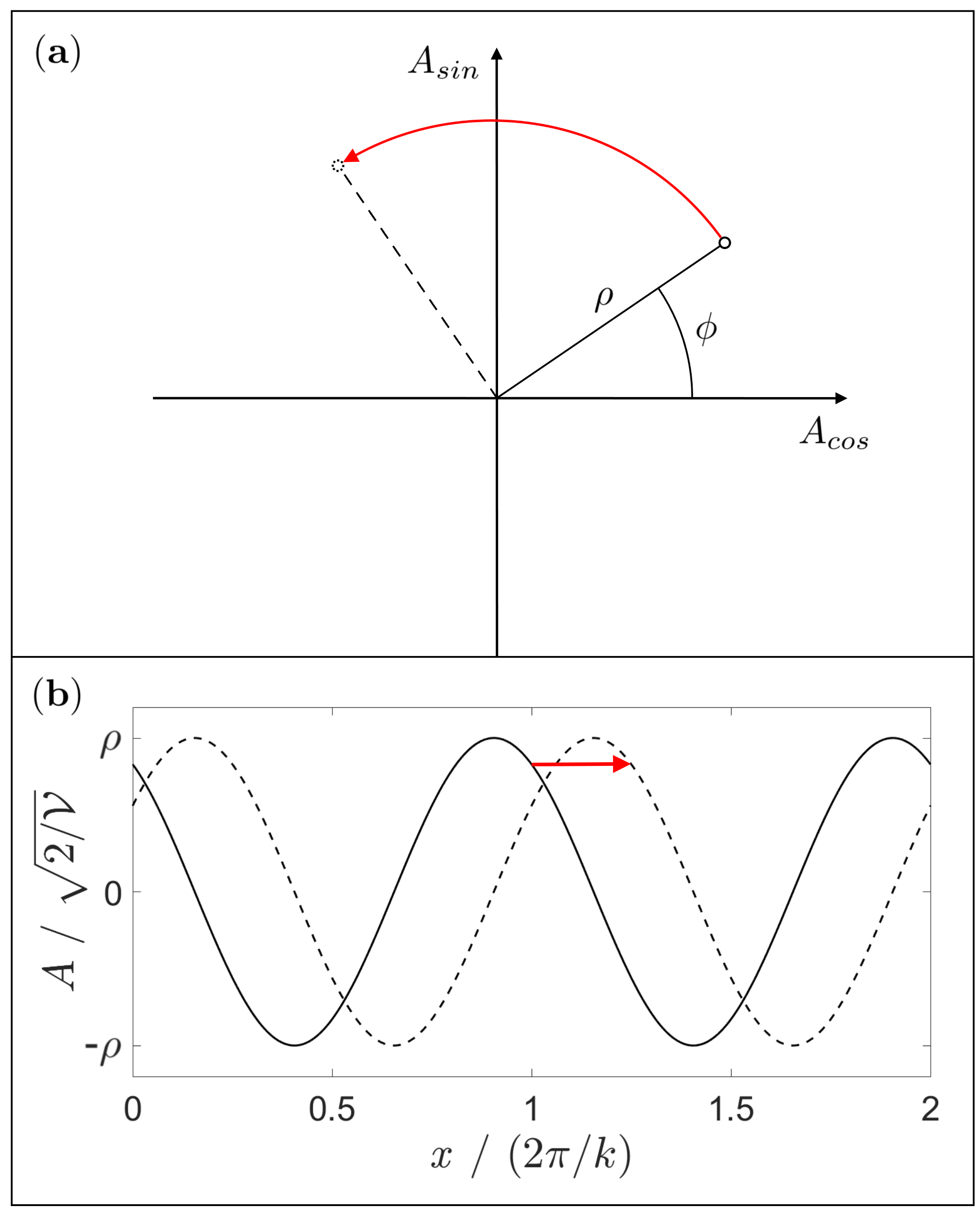}
		\caption{(a) A point with polar coordinates $(\rho, \phi)$ drawn in the $(A_{cos}, A_{sin})$-plane. The red, arching arrow illustrates a rotation of the wave function on the plane, sending the amplitude at $(\rho, \phi)$ onto another point in the plane. (b) The amplitude, $A=\bm A_{\pm\bm k, p} \cdot \bm e_{\bm k, p}$, plotted as a function of $x = \bm x \cdot \bm k/k$. 
		The solid curve corresponds to the configuration of the point $(\rho, \phi)$ in (a), and the dashed curve corresponds to the translated configuration as a result of the rotation shown in (a).
			\label{fig:2d_cossin}}
	\end{minipage}
	\hspace{0.3cm} 
	\begin{minipage}[t]{0.499\textwidth}
		\includegraphics[width=1.2 \textwidth, trim={1.3cm, 2.2cm, 0cm, 0cm}, clip]{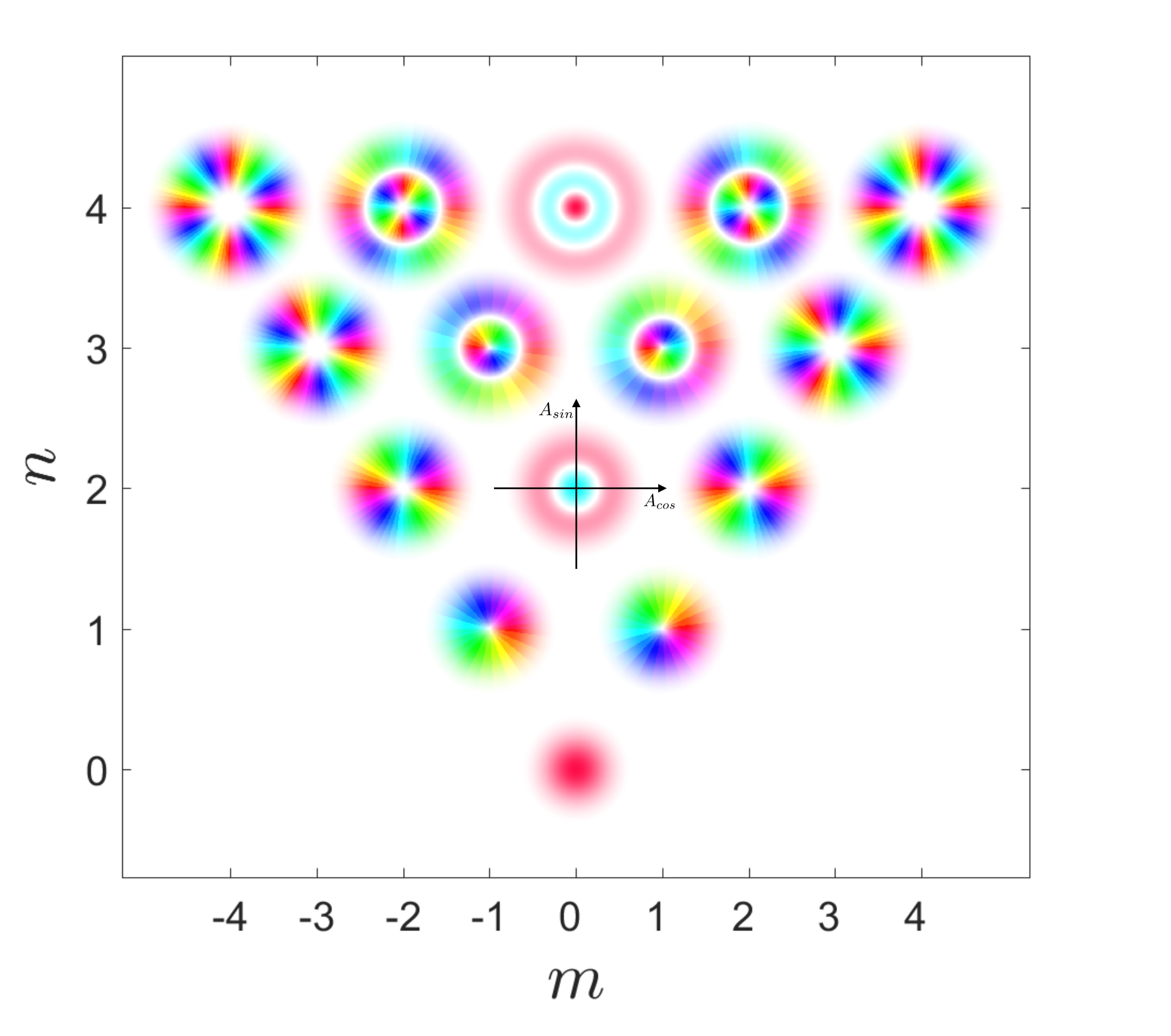}
		
		\hspace{0.32cm}
		\begin{minipage}[r]{0.99\textwidth}
			
		\vspace{0.145cm}
		\caption{The first 15 of the photon wave functions, each plotted individually in the $(A_{cos}, A_{sin})$-plane and ordered in an $m, n$-coordinate system, where $n = n_{\bm k} + n_{-\bm k}$ and $m = n_{\bm k} - n_{-\bm k}$. The $A_{cos}$-axis and $A_{sin}$-axis are illustrated only for the $(m=0, n=2)$-state. The colors represent the phase of the wave functions, which runs from $0$ to $2\pi$ as the colors run through the rainbow colors from red to violet and back to red again. The color intensity represents the absolute value of the wave functions, normalized to have the same maxima and each with an offset of being completely white under 5 \% of maximum.
			\label{fig:SHO_sols}}
		\end{minipage}
    \end{minipage}
\end{figure}

Let us consider the states created by $\hat a_{\bm k}^\dagger$ and $\hat a_{-\bm k}^\dagger$. These can be written as
\begin{align}
\begin{aligned}
\ket{n_{\bm k}, n_{-\bm k}} = \frac{1}{\sqrt{n_{\bm k}! \, n_{-\bm k}!}} (\hat a_{\bm k}^\dagger)^{n_{\bm k}} (\hat a_{-\bm k}^\dagger)^{n_{-\bm k}} \ket{0},
\label{ket_n_dg}
\end{aligned}
\end{align}
where $n_{\bm k}$ and $n_{-\bm k}$ are non-negative integers denoting the numbers of photons with wave vector $\bm k$ and $-\bm k$ respectively.
When viewed in the $(A_{cos}, A_{sin})$-plane, these states can be identified\footnote{For a more thorough treatment of these angular momentum states of the two-dimensional harmonic oscillator, see Cohen-Tannoudji et al.\ \cite{Cohen-Tannoudji et al.}.} as the simultaneous eigenstates of $H_{\pm \bm k, p}$ and the generator of rotations in the plane, $\hat L$, which is given by
\begin{equation}
\hat L = \hat A_{cos} \hat p_{sin} - \hat A_{sin} \hat p_{cos} =
i \big(\hat a_{cos} \hat a_{sin}^\dagger - \hat a_{cos}^\dagger \hat a_{sin}\big).
\label{L1}
\end{equation}
This fact can be seen by using relations (\ref{N_d}) and (\ref{N_g}) to rewrite $\hat L$ in terms of number operators as
\begin{equation}
\hat L = \hat a_{\bm k}^\dagger \hat a_{\bm k} - \hat a_{-\bm k}^\dagger \hat a_{-\bm k}.
\label{L2}
\end{equation}
It follows that
\begin{equation}
\hat L \ket{n_{\bm k}, n_{-\bm k}} = (n_{\bm k} - n_{-\bm k})
\ket{n_{\bm k}, n_{-\bm k}},
\label{L_eig}
\end{equation}
which shows that $\ket{n_{\bm k}, n_{-\bm k}}$ is an eigenstate of $\hat L$.

The observable associated with $\hat L$ is normally referred to as the ``angular momentum" of the wave function. We will stick with this name in this paper, but this angular momentum of the wave function must not be confused with the actual, physical angular momentum of the photons. Remember that we are only looking at photons with a single, linear polarization, $p$, which have no angular momentum by themselves. In fact, we see from eg.\ (\ref{L_eig}) that $\hat L$ is instead proportional to the combined momentum of the photon states with proportionality constant $1/k$, such that
\begin{equation}
\hat p_{\pm \bm k}= k \hat L.
\label{p_kL}
\end{equation}
Here, $\hat p_{\pm \bm k}$ is the momentum operator on the subspace of the $\ket{n_{\bm k}, n_{-\bm k}}$ states, such that the full photon momentum operator of the total $\bm k$-space, $\hat p$, is given by
\begin{equation}
\hat p = \sum_{k_1 > 0} \hat p_{\pm \bm k}.
\end{equation}
The relationship between $\hat L$ and $\hat p$ is in accordance with the fact that a rotation of the wave function in the $(A_{cos}, A_{sin})$-plane by an angle, $\theta$, corresponds to a translation of the $\bm A_{\pm \bm k, p}$ field by a distance of $\theta/k$, as illustrated in fig.\ \ref{fig:2d_cossin}. Since $\hat p_{\pm \bm k}$ is the generator of translations of $\bm A_{\pm \bm k, p}$, this is equivalent of stating that
\begin{align}
\begin{aligned}
e^{-i \frac{\theta}{k} \hat p_{\bm k}} = e^{-i \theta \hat L},
\end{aligned}
\end{align}
which is indeed true according to eq.\ (\ref{p_kL}).

The wave functions of the $\ket{n_{\bm k}, n_{-\bm k}}$ states, call them $\psi_{n_{\bm k}, n_{-\bm k}}$, can be calculated from eq.\ (\ref{ket_n_dg}), using the relations for $\hat a_{\bm k}$ and $\hat a_{-\bm k}$ given in eq.\ (\ref{a_d}) and (\ref{a_g}).
Cohen-Tannoudji et al.\ \cite{Cohen-Tannoudji et al.} give the following results for the first six wave functions, $\psi_{n_{\bm k}, n_{-\bm k}}$, written in polar coordinates:
\begin{align}
\begin{aligned}
n &= 0 \quad
\quad \hspace{0.35em}
m = 0 \quad \quad \quad \hspace{0.85em}
\psi_{0, 0}(\rho, \phi) = \dfrac{\sqrt{k}}{\sqrt{\pi}} e^{-\frac{k}{2} \rho^2},
\vspace{0.18cm} \\
n &= 1 \quad \hspace{0.09em}
\begin{cases}
\vphantom{\bigg)}\\
\vphantom{\bigg)}
\end{cases} \mkern-18mu
\begin{matrix*}[l]
m = 1 \quad \quad &\psi_{1, 0}(\rho, \phi) = \dfrac{\sqrt{k}}{\sqrt{\pi}} \sqrt{k}\rho e^{-\frac{k}{2} \rho^2} e^{i \phi},
\vspace{0.18cm} \\
m = -1 \quad \quad &\psi_{0, 1}(\rho, \phi) = \dfrac{\sqrt{k}}{\sqrt{\pi}} \sqrt{k}\rho e^{-\frac{k}{2} \rho^2} e^{-i \phi},
\vspace{0.18cm}
\end{matrix*} \\
n &= 2 \quad
\begin{cases}
\vphantom{\bigg)}\\
\vphantom{\bigg)}\\
\vphantom{\bigg)}
\end{cases} \mkern-16mu
\begin{matrix*}[l]
m = 2 \quad \quad &\psi_{2, 0}(\rho, \phi) = \dfrac{\sqrt{k}}{\sqrt{2 \pi}}(\sqrt{k}\rho)^2 e^{-\frac{k}{2} \rho^2} e^{2 i \phi},
\vspace{0.18cm} \\
m = 0 \quad \quad &\psi_{1, 1}(\rho, \phi) = \dfrac{\sqrt{k}}{\sqrt{\pi}}\big((\sqrt{k}\rho)^2-1\big)
e^{-\frac{k}{2} \rho^2},
\vspace{0.18cm} \\
m = -2 \quad \quad &\psi_{0, 2}(\rho, \phi) = \dfrac{\sqrt{k}}{\sqrt{2 \pi}}(\sqrt{k}\rho)^2 e^{-\frac{k}{2} \rho^2} e^{-2 i \phi}.
\vspace{0.18cm}
\end{matrix*}
\label{polar1}
\end{aligned}
\end{align}
Here, $n = n_{\bm k} + n_{-\bm k}$ is the overall occupation number, and $m = n_{\bm k} - n_{-\bm k}$ is the eigenvalue of $\hat L$, i.e.\ the overall momentum divided by $k$.
Whether $\psi$ is given in polar or Cartesian coordinates in the following will be clear from the context.

Figure (\ref{fig:SHO_sols}) shows the first 15 photon wave functions, each plotted individually in the $(A_{cos}, A_{sin})$-plane and ordered next to each other in an $m, n$-coordinate system.

As a final point in this section, let us define the complex amplitude operator, $\hat{\mathcal{A}}$, by
\begin{align}
\hat{\mathcal{A}} = \frac{1}{\sqrt{2}} \big(\hat A_{cos} - i \hat A_{sin} \big).
\label{scriptA1}
\end{align}
By using eq.\ (\ref{Ap_cossin}) as well as eq.\ (\ref{a_d}) and (\ref{a_g}), we can rewrite $\hat{\mathcal{A}}$ in terms of $\hat a_{\bm k}$ and $\hat a_{-\bm k}$ as
\begin{align}
\hat{\mathcal{A}} = \frac{1}{2\sqrt{k}} \big(\hat a_{cos} + \hat a_{cos}^\dagger -i \hat a_{sin} -i \hat a_{sin}^\dagger\big)
= \frac{1}{\sqrt{2 k}} \big(\hat a_{\bm k} + \hat a_{-\bm k}^\dagger \big).
\label{scriptA2}
\end{align}
If we label this $\hat{\mathcal{A}}$ by $\bm k$ and $p$, we see from eq.\ (\ref{A1}) that $\hat{\bm A}(\bm x)$ can thus be written as
\begin{align}
\hat{\bm A}(\bm x) =
\frac{1}{\sqrt{\mathcal{V}}} \sum_{\bm k, p}
\hat{\mathcal{A}}_{\bm k, p}\, e^{i \bm k \cdot \bm x} \bm e_{\bm k, p},
\label{A3}
\end{align}
or, since $\hat{\mathcal{A}}_{-\bm k, p} = \hat{\mathcal{A}}_{\bm k, p}^\dagger$,
\begin{align}
\hat{\bm A}(\bm x) =
\frac{1}{\sqrt{\mathcal{V}}} \sum_{k_1 > 0, p}
\big(\hat{\mathcal{A}}_{\bm k, p}\, e^{i \bm k \cdot \bm x} + \hat{\mathcal{A}}_{\bm k, p}^\dagger\, e^{-i \bm k \cdot \bm x}\big) \bm e_{\bm k, p}.
\label{A4}
\end{align}
Note that $\hat{\bm A}(\bm x)$ in terms of these equations, as well as eq.\ (\ref{A2}), is exactly like in the classical picture, just with hats put on the coefficients to mark them as operators.

\section*{The phase operator}
Having transformed $\hat{\bm A}$ in terms of cosine and sine functions, it is now easy to identify the phase operator, $\hat E$. If we choose $\hat E$ to represent the complex observable $\exp(-i \phi)$, such that
\begin{equation}
\hat E = e^{-i \hat \phi},
\label{E0}
\end{equation}
then for all $\psi$ we simply have
\begin{equation}
(\hat E \, \psi) (A_{cos}, A_{sin})  =
\bigg(\frac{A_{cos}}{\sqrt{A_{cos}^2+A_{sin}^2}} - i \frac{A_{sin}}{\sqrt{A_{cos}^2+A_{sin}^2}}
\bigg) \psi(A_{cos}, A_{sin}).
\label{phase1}
\end{equation}
The fact that the denominator vanishes in $(0, 0)$ is not a problem since one can always remove any finite number of points from the domain of the wave functions without changing the Hilbert space.\footnote{See for example Durhuus and Solovej \cite{Durhuus and Solovej}, if not other literature on Hilbert spaces of square-integrable functions.} The reason for this is connected with the fact that any vector in a Hilbert space must be uniquely determined by its inner products with all vectors in a complete basis. Since removing any point from the domain does not change any inner products on the space, the Hilbert space therefore remains unchanged. We can thus omit $(0, 0)$ from the $(A_{cos}, A_{sin})$-plane. Note that removing a point from the domain does not change the differentiability of the wave functions either.

Having omitted $(0, 0)$ from the $(A_{cos}, A_{sin})$-plane, we see that $\hat{\mathcal{A}} \hat{\mathcal{A}}^\dagger =(\hat A_{cos}^2+\hat A_{sin}^2)/2\,$ is now a strictly positive multiplicator on the plane. It is therefore evident that $\hat{\mathcal{A}} \hat{\mathcal{A}}^\dagger$ is positive definite, and it follows that we can take both the square root and the inverse of the operator. This means that $\hat E$ can now be written in terms of $\hat a_{\bm k}$ and $\hat a_{-\bm k}$ as
\begin{align}
\begin{aligned}
\hat E = 
\big(\hat{\mathcal{A}} \hat{\mathcal{A}}^\dagger\big)^{-\frac{1}{2}}
\hat{\mathcal{A}}=
\big((\hat a_{\bm k} + \hat a_{-\bm k}^\dagger ) (\hat a_{\bm k}^\dagger + \hat a_{-\bm k})\big)^{-\frac{1}{2}}
\big(\hat a_{\bm k} + \hat a_{-\bm k}^\dagger \big).
\label{phase2}
\end{aligned}
\end{align}
It is remarkable how similar this expression looks to the SG operator of eq.\ (\ref{SG}). The only difference is that here, $\hat a$ is replaced with $(\hat a_{\bm k} + \hat a_{-\bm k}^\dagger)$ instead. Whereas $\hat a$ in eq.\ (\ref{SG}) had the effect of annihilating a photon, the effect of the replacement, $(\hat a_{\bm k} + \hat a_{-\bm k}^\dagger)$, is now a superposition of the effects of annihilating a photon moving in one direction (the forward direction) and creating a photon moving in the opposite direction. Since the equation was derived from eq.\ (\ref{phase1}), we see that this adjustment to the SG operator has the desired effect of making it unitary.

Since $(\hat{\mathcal{A}} \hat{\mathcal{A}}^\dagger)^{-1/2}=\sqrt{2}(\hat A_{cos}^2+\hat A_{sin}^2)^{-1/2}$ commutes with $\hat L$, it can be seen from eq.\ (\ref{phase2}) that $\hat E$ can only lower the eigenvalue of $\hat L$ by $1$ for any of its eigenvectors. This fact can also be seen by considering the $\phi$-dependency of the wave function for any eigenvector of $\hat L$, which is contained in the factor $\exp(i (n_{\bm k} - n_{-\bm k}) \phi)$. Since the operator $\hat E = \exp(-i \hat \phi)$ will send this into $\exp(i (n_{\bm k} - n_{-\bm k} - 1) \phi)$, the eigenvalue of $\hat L$ will indeed be lowered by exactly $1$. This means that $\hat E \ket{n_{\bm k}, n_{-\bm k}}$ will be orthogonal to any eigenvector, $\ket{n_{\bm k}', n_{-\bm k}'}$, for which $n_{\bm k}' - n_{-\bm k}' \neq n_{\bm k} - n_{-\bm k} - 1$.
We can use this information to write $\hat E$ as
\begin{align}
\begin{aligned}
\hat E &= \sum_{n_{\bm k}=0}^{\infty}\sum_{n_{-\bm k}=0}^{\infty} \sum_{n'_{\bm k}=0}^{\infty}\sum_{n'_{-\bm k}=0}^{\infty}
\delta_{n'_{\bm k}-n'_{-\bm k}, \, n_{\bm k}-n_{-\bm k}-1}
\ket{n'_{\bm k}, n'_{-\bm k}} \braket{n'_{\bm k}, n'_{-\bm k}|
\hat E|n_{\bm k}, n_{-\bm k}}\bra{n_{\bm k}, n_{-\bm k}}.
\label{phase5}
\end{aligned}
\end{align}
Note that every matrix element, $\braket{n'_{\bm k}, n'_{-\bm k}|\hat E|n_{\bm k}, n_{-\bm k}}$, can be calculated once the wave functions of the two relevant states have been obtained, which can be done via eq.\ (\ref{ket_n_dg}).

We can now ask ourselves the interesting question of what this new phase operator will become when it is restricted to a space containing only photons moving in the forward direction and no backward moving photons. This is the space on which previous phase operators such as the SG operator have been defined. This simplified operator, call it $\hat E_+$, can be of interest when considering a beam of photons rather than a cavity mode. We will define $\hat E_+$ by
\begin{align}
\begin{aligned}
\hat E_+ = \sum_{n_{\bm k}=0}^{\infty}
\ket{n_{\bm k}, 0} \braket{n_{\bm k}, 0| \hat E |n_{\bm k}+1, 0}\bra{n_{\bm k}+1, 0},
\label{E_+1}
\end{aligned}
\end{align}
such that
\begin{align}
\begin{aligned}
\big \langle n'_{\bm k}, 0 \big| \hat E_+ \big| n_{\bm k}, 0 \big\rangle= \big \langle n'_{\bm k}, 0 \big| \hat E \big| n_{\bm k}, 0 \big\rangle
\label{E_+2}
\end{aligned}
\end{align}
for all $n_{\bm k}, n'_{\bm k} \in\lbrace 0, 1, 2, \ldots\rbrace$. If we are interested in the expectation value of $\hat E$ on this particular subspace, we can therefore just as well use $\hat E_+$ instead of $\hat E$.

Let us analyze this operator, $\hat E_+$. We can start by noting that it is not unitary. This can be shown by calculating one of the non-zero matrix elements of $\hat E$ that are set to zero in $\hat E_+$. We can for instance calculate the matrix element $\braket{1, 1|\hat E|1, 0}$. Using eq.\ (\ref{polar1}), we get
\begin{align}
\begin{aligned}
\braket{1, 1|\hat E |1, 0} &=
\frac{k \sqrt{k}}{\pi}\int_{0}^{\infty}\int_{0}^{2 \pi} (k \rho^2 - 1) \rho e^{-k \rho^2} e^{-i \phi} e^{i \phi} \rho \, d \phi \, d \rho \\
&= 2 k \sqrt{k}\int_{0}^{\infty}(k \rho^4 - \rho^2) e^{-k \rho^2}  \, d \rho \\
&= 2\int_{0}^{\infty}(x^4 - x^2)  e^{-x^2}  \, d x \\
&= \frac{\sqrt{\pi}}{4} \neq 0,
\end{aligned}
\end{align}
Now, since
\begin{align}
\begin{aligned}
\hat E_+ \ket{1, 0} = \hat E \ket{1, 0} - \braket{1, 1| \hat E | 1, 0}\ket{1, 1}
- \braket{2, 2| \hat E | 1, 0}\ket{2, 2} - \ldots,
\end{aligned}
\end{align}
we see that
\begin{align}
\begin{aligned}
\braket{1, 0|\hat E_+^\dagger \hat E_+| 1, 0}=\braket{1, 0|\hat E^\dagger \hat E| 1, 0}
- |\braket{1, 1| \hat E | 1, 0}|^2 - \ldots = 1 -\frac{\pi}{16} - \ldots < 1.
\end{aligned}
\end{align}
We can therefore conclude that $\hat E_+$ is indeed not unitary. This fact might help us understand why no one has succeeded in finding an appropriate unitary phase operator when looking at photons with only a particular wave vector $\bm k$, and not at the same time including the photons with wave vector $-\bm k$. The phase operator is simply not supposed to be unitary when restricted to this space.

Let us proceed to express $\hat E_+$ in terms of its matrix elements. It turns out that there is a very neat formula for these. To get them, we will use the fact that, according to Cohen-Tannoudji et al.\ \cite{Cohen-Tannoudji et al.}, the general wave function for $\ket{n_{\bm k}, 0}$ is given in polar coordinates by
\begin{align}
\begin{aligned}
\psi_{n_{\bm k}, 0}(\rho, \phi) = \frac{\sqrt{k}}{\sqrt{\pi}\sqrt{n_{\bm k} !}\,}
(\sqrt{k} \rho)^{n_{\bm k}} e^{-\frac{k}{2} \rho^2} e^{i n_{\bm k} \phi}.
\end{aligned}
\end{align}
The matrix element $\braket{n_{\bm k}, 0|\hat E|n_{\bm k} + 1, 0}$ is therefore given by
\begin{align}
\begin{aligned}
\braket{n_{\bm k}, 0|\hat E |n_{\bm k} + 1, 0} &=
\frac{k}{\pi \sqrt{n_{\bm k}! (n_{\bm k}+1)!}}
\int_{0}^{\infty}\int_{0}^{2 \pi}
(\sqrt{k} \rho)^{2 n_{\bm k}+1} e^{-k \rho^2}
\rho \, d \phi \, d \rho
\\ &=
\frac{2 k}{n_{\bm k}! \sqrt{n_{\bm k}+1}}
\int_{0}^{\infty}
(\sqrt{k} \rho)^{2 n_{\bm k}+1} e^{-k \rho^2}
\rho \, d \rho
\\ &=
\frac{1}{n_{\bm k}! \sqrt{n_{\bm k}+1}}
\int_{0}^{\infty}
x^{n_{\bm k}+\frac{1}{2}} e^{-x}
\, d x
\\ &=
\frac{1}{n_{\bm k}! \sqrt{n_{\bm k}+1}}
\Gamma \Big(n_{\bm k} + 1 + \frac{1}{2} \Big),
\end{aligned}
\end{align}
where we have changed the integration variable to $x=k \rho^2$ to get the third equality and then identified the resulting integral as the gamma function. If we write $\Gamma (n_{\bm k} + 1 + 1/2)$ more compactly as $(n_{\bm k}+1/2)!$, we see that eq.\ (\ref{E_+1}) now becomes
\begin{align}
\begin{aligned}
\hat E_+ = \sum_{n_{\bm k}=0}^{\infty}
\frac{(n_{\bm k}+\frac{1}{2})!}{n_{\bm k}! \sqrt{n_{\bm k}+1}}
|n_{\bm k}, 0 \rangle \langle n_{\bm k}+1, 0 |.
\label{E_+3}
\end{aligned}
\end{align}
The sequence
\begin{align}
\begin{aligned}
(a_n)_{n \in \lbrace 0, 1, 2, \ldots \rbrace} =
\Big ( \frac{(n+\frac{1}{2})!}{n! \sqrt{n+1}} \Big)_{n \in \lbrace 0, 1, 2, \ldots\rbrace} \approx
(0.8862, 0.9400, 0.9594, 0.9693, 0.9754, \ldots)
\end{aligned}
\end{align}
is convergent, and it converges to $1$. A way to see this is to use the duplication formula:
\begin{align}
\begin{aligned}
\Gamma(n+1)\Gamma(n+1+\frac{1}{2}) = 2^{-2 n-1} \sqrt{\pi} \, \Gamma(2n + 2)
\Rightarrow
\frac{(n+\frac{1}{2})!}{n!} = 2^{-2 n-1} \sqrt{\pi}  \frac{(2 n + 1) (2 n)!}{(n!)^2}.
\end{aligned}
\end{align}
Stirling's approximation tells us that
\begin{align}
\begin{aligned}
\frac{(2 n)!}{(n!)^2} \sim \frac{\sqrt{4 \pi n} e^{-2 n} (2 n)^{2 n}}
{2 \pi n e^{-2 n} n^{2 n}}
= \frac{4^n}{\sqrt{\pi n}}
\end{aligned}
\end{align}
for $n$ tending toward infinity, and therefore we have
\begin{align}
\begin{aligned}
\lim_{n\to \infty} a_n = 
\lim_{n\to \infty} \frac{1}{\sqrt{n+1}} \frac{1}{2}\frac{1}{4^n}\sqrt{\pi} (2 n + 1)
\frac{4^n}{\sqrt{\pi n}}
=
\lim_{n\to \infty} \frac{2 n + 1}{2 \sqrt{n} \sqrt{n+1}}
= 1.
\end{aligned}
\end{align}
To show that $a_n<1$ for all $n$, one can quite easily calculate $a_{n+1}-a_n$ to show that $a_{n+1}-a_n>0$ for all $n$. The strictly increasing sequence will therefore never reach 1.

The similarity between our $\hat E_+$ and the SG operator, which is also defined on the space of states with only forward moving photons, is now even more striking. If we rename all $\ket{n_{\bm k}, 0}$ as $\ket{n}$, $n=n_{\bm k} \in \lbrace0, 1, 2, \ldots \rbrace$, for compactness, we can write $\hat E_+$ as
\begin{align}
\begin{aligned}
\hat E_+ = \sum_{n=0}^{\infty}
a_n
|n \rangle \langle n+1|.
\label{E_+4}
\end{aligned}
\end{align}
Meanwhile, the SG operator, as we recall, is given by
\begin{align}
\begin{aligned}
\hat E_{SG} = \sum_{n=0}^{\infty}
|n \rangle \langle n+1|.
%\label{SG2}
\end{aligned}
\end{align}
This shows that the SG operator is a very good approximation to $\hat E_+$, especially for large occupation numbers. We just need to remember that $\hat E_+$ represents $\exp(- i \hat \phi)$ in this paper, and not $\exp(i \hat \phi)$ like it does, for instance, in Gerry and Knight \cite{Gerry and Knight}.

A natural question we can ask ourselves moving on is what the uncertainty is on a measurement of the phase factor of a beam of photons. Since $\hat E_+$ is not hermitian, we cannot use the commutator to get a lower limit on e.g.\ $\Delta N \Delta E_+$. We can, however, still calculate the variance of $ \hat E$ for any state in the full Hilbert space. It is given by
\begin{align}
\begin{aligned}
|\Delta E|^2 &=
\big\langle\big(e^{i \hat \phi}-\langle\hat e^{i \hat \phi}\rangle\big)\big(e^{- i \hat \phi}-\braket{e^{- i \hat \phi}}\big)\big\rangle \\ &=
\big\langle\big(\hat E^\dagger-\langle\hat E\rangle^*\big)\big(\hat E-\braket{\hat E}\big)\big\rangle
\\ &=
\braket{\hat E^\dagger \hat E} - |\langle\hat E\rangle|^2
\\ &=
1 - |\langle\hat E\rangle|^2.
\label{Delta_E1}
\end{aligned}
\end{align}
When we are interested in a state with only forward moving photons, $\ket{\psi}=\sum_{n=0}^{\infty} c_n \ket{n}$, we can thus use eq.\ (\ref{E_+2}) and (\ref{E_+4}) to obtain
\begin{align}
\begin{aligned}
|\Delta E_+|^2 =
1 - |\langle \psi|\hat E_+ | \psi\rangle|^2 =
1 - \big|\sum_{n=0}^{\infty} a_n c_n^* c_{n+1}\big|^2.
\label{Delta_E2}
\end{aligned}
\end{align}
To check that the RHS of this formula is greater than $0$, define two new vectors: $\ket{\psi_1}=\sum_{n=0}^{\infty} a_n c_n \ket{n}$ and $\ket{\psi_2}=\sum_{n=0}^{\infty} c_{n+1} \ket{n}$. Note that $\braket{\psi_1|\psi_1} < 1 \geq \braket{\psi_2|\psi_2}$, and therefore, due to the Cauchy-Schwarz inequality,
\begin{align}
\begin{aligned}
|\Delta E_+|^2 = 1 - |\braket{\psi_1 | \psi_2}|^2 > 0. 
\end{aligned}
\end{align}

Following up on this, we can use eq.\ (\ref{Delta_E2}) to try searching for a state, $\ket{\psi}$, that minimizes $|\Delta E_+|^2$. If we define a state, $\ket{\psi}$, by
\begin{align}
\begin{aligned}
\ket{\psi}=\sum_{n=l}^{l+m-1} \frac{1}{\sqrt{m}} \ket{n},
\end{aligned}
\end{align}
we have, for large enough $l$,
\begin{align}
\begin{aligned}
|\Delta E_+|^2 =
1 - \big|\sum_{n=0}^{\infty} a_n c_n^* c_{n+1}\big|^2 \approx 1 - \big|\sum_{n=0}^{\infty} c_n^* c_{n+1}\big|^2 = 1 -\frac{(m-1)^2}{m^2} = \dfrac{2m-1}{m^2}.
\end{aligned}
\end{align}
This result goes to $0$ as $m$ goes to infinity, and we can therefore get an arbitrarily small $|\Delta E_+|^2$ by choosing $m$ large enough. This shows that we can still get an arbitrarily precise phase for a photon beam even without any backward moving photons.
The time evolution of this $\ket{\psi}$ is given by
\begin{align}
\begin{aligned}
\ket{\psi(t)}=\sum_{n=l}^{l+m-1} \frac{e^{-i n k t}}{\sqrt{m}} \ket{n},
\end{aligned}
\end{align}
and therefore the time evolution of the expectation value of $\hat E_+$ is
\begin{align}
\begin{aligned}
\braket{\psi(t) | \hat E_+ | \psi(t)} = \sum_{n=l}^{l+m-1}
a_n  \frac{e^{i n k t}}{\sqrt{m}}  \frac{e^{-i (n + 1) k t}}{\sqrt{m}}
\approx \frac{m - 1}{m} e^{- i k t}
\end{aligned}
\end{align}
for large enough $l$. The result tends toward $\exp(-i k t)$ as $m$ tends to infinity, which is as expected for a coherent photon beam.

As a final remark in this paper, I will point out that it might be beneficial to look at operators $\hat{\mathcal{A}}$ or $\hat{\mathcal{A}} \hat{\mathcal{A}}^\dagger$ instead of $\hat E$ when analyzing the time evolution or the uncertainty of the $\bm A$ field, especially when it contains both backward and forward moving photons. This should make the calculation easier since these operators contain only positive, integer powers of the ladder operators. If, for instance, one wants to find the behavior of the $\bm A$ field of a coherent cavity mode, one might therefore choose to look at the expectation value of $\hat{\mathcal{A}}$.

\section*{Conclusion}
We have derived the photon phase operator starting from just the well known vector potential operator, $\hat{\bm A}$, and the well known Hamiltonian for photons. As it turned out, only a simple transformation was needed to make $\hat{\bm A}$ exactly analogous to the expression for the classical $\bm A$ field. From there the phase operator was derived easily. The result was similar to the well known SG operator, but defined on a larger space with photons having wave vectors equal to both $\bm k$ and $-\bm k$. It seems that the reason why no one has found a satisfactory, unitary phase operator so far might be because no one has tried including the backward moving photons in the domain of the operator and tried to combine both $\hat a_{\bm k}$ and $\hat a_{-\bm k}$ in the definition. 
When restricted to the space of photons with wave vectors only equal to $\bm k$, setting all other matrix elements to zero, the phase operator turned out to be numerically very close to the SG operator as well. The correction is a factor of $a_n = (n+1/2)!/(n! \sqrt{n+1})$ on the $n$th matrix element for all $n$, and this correction becomes less significant with increasing photon number.

\end{document}